# JURNAL EKONOMI KUANTITATIF TERAPAN

# The Impact of COVID-19 on FinTech Lending in Indonesia: Evidence From Interrupted Time Series Analysis

Abdul Khaliq

**ABSTRAK**

*Penelitian ini mengukur dampak pandemi COVID-19 terhadap financial technology (FinTech) lending di Indonesia. Menggunakan data bulanan FinTech yang dipublikasikan oleh Otoritas Jasa Keuangan pada periode 2018M02-2021M04, artikel ini mengukur dampak COVID-19 yang dimulai pada tanggal 2 Maret 2020 terhadap FinTech dengan mengadopsi interrupted time series (ITS) experiment. Estimasi memperlihatkan COVID-19 memiliki dampak negatif pada perubahan level (changes in level) FinTech lending di Indonesia, tetapi perubahan tren (changes in trend) adalah positif. Selanjutnya, COVID-19 berdampak negatif dan secara statistik signifikan terhadap perubahan level Rasio Pinjaman Lancar (s.d. 90 hari). Namun, COVID-19 memberikan efek positif dan secara statistik signifikan pada perubahan level Rasio Pinjaman Macet (>90 hari). Temuan ini merekomendasikan bahwa otoritas jasa keuangan secara intensif mendorong berbagai model baru bisnis FinTech yang inovatif post-COVID19 dalam upaya memperluas inklusi keuangan digital dengan menyediakan pembiayaan bagi masyarakat (P2P) yang tidak tersentuh oleh bank.*

***Kata Kunci****: COVID-19, FinTech, Interrupted Time Series*
***Klasifikasi JEL****: I15, E42, C22*

**ABSTRACT**

This study measures the impact of COVID-19 outbreaks on financial technology (FinTech) lending in Indonesia. Using monthly FinTech data published by Financial Services Authority (OJK) over the period 2018M02-2021M04, the article examines the impact of COVID-19 started on March 2020 on FinTech by adopting an interrupted time series (ITS) experiment. The estimation shows that the COVID-19 outbreaks negatively affect changes in FinTech lending level in Indonesia, but the changes in the trend are positive. Moreover, the COVID-19 has been found to have a negative and statistically significant effect on the 90-day success loan settlement rate level. However, COVID-19 has positive and statistically significant effects on the 90-day default rate of loan repayment level. These estimation results recommend that the financial services authority of Indonesia should intensively promote various innovative financial technology (FinTech) lending post-COVID-19 to increase digital financial inclusion by providing peer to peer lending (P2P) to unbanked populations.

## INTRODUCTION

Financial technology, known as FinTech, has developed in recent years. Fintech is a combination of financial services with technology. FinTech began to grow and become widely known after the 2007-2008 financial crisis. FinTech's role as financial services provider is expected to augment during the coronavirus disease (COVID-19) pandemic as an innovative, less expensive, efficient, and reaching every side of the economy.

The COVID-19 pandemic has brought about changes to current economic activities and business models. On the one hand, the implementation of strict health protocols to reduce the spread of COVID-19 has accelerated the operation of digitalization of everything from education to economic and business activities. But, on the other hand, COVID-19 has decreased economic and business activities that lead to financial difficulties for both economic and business actors and society. In this position, FinTech lending can cut the spread of COVID-19 and provide financial solutions for financially vulnerable people.

Eventhough COVID-19 has encouraged the acceleration of the application of digital technology to all economic and business activities, COVID-19 presents challenges for the development of FinTech lending in Indonesia. The COVID-19 has had a non uniform impact on FinTech lending in Indonesia (Otoritas Jasa Keuangan, 2021). The announcement of COVID-19 in March 2020 has decreased FinTech lending activity. Figure 1 shows the evolution of FinTech lending before and during COVID-19. Before March 2020, FinTech lending increased, but starting in April 2020, FinTech lending decreased. The total number of FinTech loans in March 2020 has reached Rp.7139.82 billion. In May 2020, the total loan amount fell to Rp.3116.07 billion.

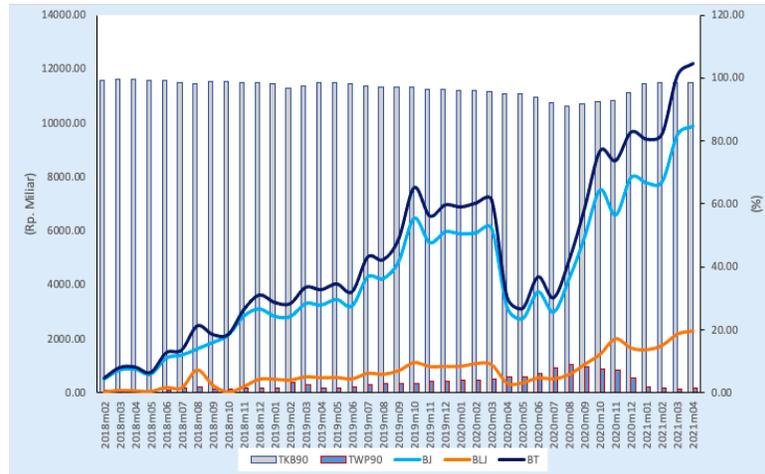

Source: Otoritas Jasa Keuangan (2021)
Figure 1 Development of Fintech Lending in Indonesia

Figure 1 shows the fluctuating FinTech lending during the COVID-19 pandemic, but the fundamental question that needs answers is how much the impact does COVID-19 has on FinTech lending in Indonesia? Study that measuring the impact of COVID-19 on FinTech lending in Indonesia has never been conducted yet. Making conclusion of the effect of COVID-19 on FinTech lending only from the illustration of FinTech lending data during the COVID-19 pandemic and comparing it with Fintech lending data before the COVID-19 pandemic is naïve with weak scientific methods. Therefore, to measure the impact of COVID-19 on FinTech lending, it is essential to reconstruct counterfactual FinTech lending data without COVID-19. The counterfactual data reconstruction requires specific experimental research methods.

This paper enriches the literature and empirical findings in measuring the impact of COVID-19 on FinTech lending in Indonesia. This information can provide researchers, policymakers, and businesses with an understanding of the

245

impact of COVID-19 on building a new platform business model of FinTech lending in Indonesia.

## LITERATURE REVIEW

In the past decade, financial technology (FinTech) has attracted global attention. The term of FinTech was originally introduced by Charnes, Raike, & Bettinger (1972). This definition is reinforced by integrating technology and finance (Gomber, Koch, & Siering, 2017). Meanwhile, Navaretti, Calzolari, Mansilla-Fernandez, & Pozzolo (2018) defines FinTech as FinTech companies and classifies it according to FinTech business types, such as FinTech payment businesses and FinTech lending businesses. Subsequent, Cheng & Qu (2020) define FinTech banks as emerging technology applications within the banking industry.

Although the definition of FinTech is still debated among academicians, the development of financial technology literature is basically based on two interrelated concepts, namely internet finance and FinTech (Cheng & Qu, 2020). According to Cheng & Qu (2020),

Internet finance is related to the combination of finance and internet technology, while FinTech combines finance and technological developments such as artificial intelligence (AI), blockchain technology, big data technology, and internet technology. Internet finance emphases on the meaning of internet finance and its characteristics (Berger & Gleisner, 2009) and the effect of internet finance on the economy and finance (Krueger, 2012). Meanwhile, the FinTech discusses the characteristics and composition of FinTech (Gomber et al., 2017; Valle & Zeng, 2019).

At this time, FinTech studies in Indonesia have undergone developments at both macro and micro level perspectives. At the macro level perspective, Siantur (2017) discusses the impact of digital technology on economic growth. Meanwhile, at the micro-level perspective, FinTech studies look more at the effect of FinTech on the development of MSMEs (Andaiyani, Yunisvita, & Tarmizi, 2020; Ningsih, 2020; Rahmawati, Rahayu, Nivanty, &

Lutfiah, 2020). In addition, FinTech provides financial solutions for micro small and medium enterprises, refer to MSMEs, (Ningsih, 2020). In addition, Rahardjo, Khairul, & Siharis (2019) state that FinTech has a role in improving the performance of MSMEs through its function as a marketplace for MSMEs production and MSMEs buying and selling.

During the COVID-19 pandemic, FinTech discussions shifted to the impact of COVID-19 on the development of FinTech, Nasution, Ramli, & Sadalia (2020) and Yudhira (2021) among others. Nasution et al. (2020) show that the COVID-19 has changed the behavior of transactions using digital finance in Indonesia. Then, Yudhira (2021) COVID-19 has opened up opportunities for Islamic FinTech due to the change in people's lifestyle to digitalization.

Although many researchers have conducted investigations on FinTech and internet finance in the past decade, studies that measure the impact of COVID-19 on FinTech are still limited. Some studies only look at the development of FinTech during the COVID-19 pandemic and compare it to FinTech's condition before Covid-19. The downside of this kind of method is naïve with biased conclusion-making. To measure the impact of COVID-19 on FinTech requires a specific methodology. Therefore, this study focuses on the impact of the COVID-19 pandemic on FinTech lending in Indonesia using the interrupted time series (ITS) analysis method.

## METHODS AND DATA

This article uses secondary FinTech lending data published by Otoritas Jasa Keuangan (2020) period 2018M02-2021M04. FinTech lending unit is in nominal billions of rupiah. Measurement of the 90-day success loan settlement rate (TKB90) and the 90-day default rate of loan repayment (TWP90) are in percentage form. The amount of FinTech lending proxy is distinguished into loan amount from Java island (BJ), loan amount from outside Java island (BLJ), and aggregate loan (BT). Meanwhile, COVID-19 information uses publication data Satuan Tugas Penanggulangan COVID-19 (2020). The COVID-19 intervention started in March 2020 when the first case of COVID-19 was officially announced by Satuan Tugas Penanggulangan COVID-19 (2020) March 2, 2020 (2020M03) or the 26th month of the 2018M01-2021M04 time frame.

In the literature, the best method of impact measurement (gold standard) is used experimental research randomized controlled trial (RCT). However, RCT has stringent criteria in design, implementation, and analysis, so alternative methods of impact measurement can refer to the use of quasi-experimental research (Hudson, Fielding, & Ramsay, 2019). The best quasi-experimental research observation method is the interrupted times series (ITS) approach (Cook, Thomas D; Campbell, 1979; Penfold & Zhang, 2013).

In general, ITS single-group regression standard model, according to Linden (2015), is following the equation:

$$Y_t = \beta_0 + \beta_1 T_t + \beta_2 X_t + \beta_3 X_t T_t + \epsilon_t, \quad (1)$$

where $Y_t$ is the outcome at the time of t, $X_t$ is a dummy variable that indicates the intervention (pre-intervention period 0 and 1 other), $T_t$ is the time t. $\beta_0$ is a base level measurement of the outcome at the beginning of the series. $\beta_1$ is the estimated base trend. $\beta_2$ is measuring changes in level after intervention (post-intervention segment). $\beta_3$ calculates the changes in trend after intervention (post-intervention segment). And $\epsilon_t$ is an error process.

Refers to ITS standard model by Ferron & Rendina-Gobioff (2015) dan Linden (2015) the visualization of the impact of COVID-19 on FinTech lending is

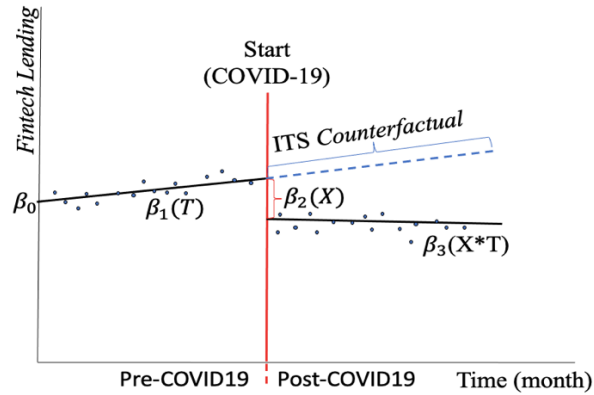

Figure 2 Visualization of ITS Standard Model

Specifically, ITS single-group regression standard model examines the impact of COVID-19 on FinTech in the form of segmented regression equations:

$$FinTech_t = \beta_0 + \beta_1 T_t + \beta_2 COVID19_t + \beta_3 COVID19_t T_t + \epsilon_t, \quad (2)$$

where $FinTech_t$ is FinTech lending at time t, $COVID19_t$ is dummy variables (indicators) that reflect COVID19 interventions (pre-COVID19 period 0 and 1 other), $T_t$ is the time t, $\beta_0$ is the estimated base level of FinTech at the beginning of the series (pre-COVID19 intercept), $\beta_1$ is the estimated base trend (pre-COVID19 slope), $\beta_2$ measures the coefficient of change in level after intervention (post-COVID19 change in intercept), $\beta_3$ estimates the coefficient of change in trend after intervention (post-COVID19 change in slope), $\epsilon_t$ is an error process. Furthermore, this equation is estimated by utilizing the generalized least square (GLS) and autoregressive moving average (ARMA) techniques utilizing the R package'nlme' (Pinheiro et al., 2020).

## RESULTS AND ANALYSIS

The estimated impact of the COVID-19 pandemic on FinTech lending in Indonesia using interrupted time series (ITS) for 2019M02-2021M04 shows diverse results. In general, COVID-19 has a negative impact on FinTech lending. Furthermore, COVID-19 has a negative effect on the 90-day success loan settlement rate and positively impacts on the 90-day default rate of loan repayment. The estimation results of the impact of COVID-19 on FinTech lending can be seen in Table 1 and Figure 2. Meanwhile, the effect of COVID-19 on the 90-day success loan settlement rate and the 90-day default rate of loan repayment are stated in Table 2 and Figure 3.

Table 1. Estimated Results of Segmented Regression Model For FinTech Lending

|  | BJ | BLJ | BT |
|---|---|---|---|
| Constant | -27.782 | -4.442 | -25.251 |
|  | (-0.134) | (-0.039) | (-0.098) |
| Time | 243.993*** | 40.581*** | 284.144*** |
|  | (16.199) | (5.259) | (15.114) |
| Post-COVID19 period | -3665.934*** | -762.727*** | -4408.780*** |
|  | (-7.958) | (-4.163) | (-7.738) |
| Time x post-COVID19 period | 248.572*** | 96.908*** | 341.676*** |
|  | (5.576) | (4.865) | (6.176) |

Note: ***, **, * are at 1%, 5%, and 10% and the numbers in parentheses are t-statistics.

Table 1 and Figure 3 show the impact of COVID-19 on FinTech lending, which explicitly tracks the number of loans from Java (BJ), the number of loans from outside Java (BLJ), and the total loan amount (BT). Before the government announced the COVID-19 pandemic on March 2, 2020, the pre-COVID19 slop of intervention on FinTech loans showed a positive and statistically significant effect at the level of 1%. The positive effect was shown by each slop the number of loans by borrower from Java with the amount of IDR 243.9 billion, the number of loans by borrower from outside Java amounted to IDR 40.58 billion, and the aggregate loan with the amount of IDR 284.14 billion.

The Impact of COVID-19 on FinTech Lending in Indonesia: Evidence…..

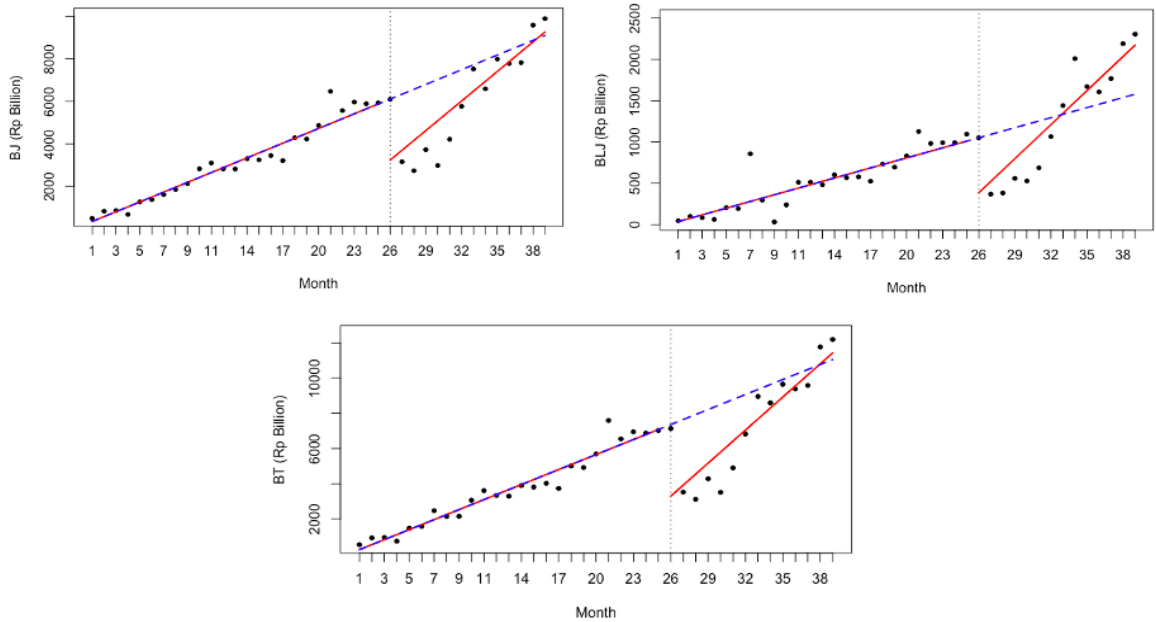

Figure 3 Segmented Regression Models For the Impact of COVID-19 on FinTech Lending Starting in the 26th Month (March 2020)

After the COVID-19 pandemic was officially announced by the government of the Republic of Indonesia on March 2, 2020, the number of new loans from FinTech decreased indicated by changes in intercept. The decrease in the number of new loans from FinTech is seen in the negatively marked intercept changes and statistically significant at the level of 1%. The estimated result shows a change in the post-COVID19 level in borrower from Java of negative IDR 3665.93 billion, the amount of loans by borrower from outside Java with a negative amount of IDR 762.73 billion, and the total number of loans decreased by IDR 4408.78 billion. Nevertheless, the trend of loans from FinTech after the announcement of COVID-19 on March 2, 2020, shows a positive slope of change in trend post-COVID19 and is statistically significant at the level of 1%.

Figure 3 shows the changes in trend of loans from FinTech for borrower from Java and borrower from outside Java. The trend of FinTech loans for pre-COVID19 interventions and post-COVID19 interventions has increased. The FinTech lending trend by borrower from Java post-COVID19

251

(March 2, 2020) has increased, but the trend line is still below the pre-COVID19 trend. In addition, FinTech loan trend from outside Java has increased, and the post-COVID19 trend line since October 2020 has been above the pre-COVID19 trend line. Total FinTech lending trend have increased with the post-COVID19 trend line, following the pre-COVID19 trend since March 2021.

Table 2. Estimated Results of Segmented Regression Model For FinTech Lending Quality

|  | TKB90 | TWP90 |
|---|---|---|
| Constant | 100.344*** | -0.336 |
|  | (167.158) | (-0.558) |
| Time | -0.195*** | 0.194*** |
|  | (-4.839) | (4.823) |
| Post-COVID19 period | -0.308 | 0.306 |
|  | (-0.545) | (0.564) |
| Time x post-COVID19 period | 0.184 | -0.183 |
|  | (1.426) | (-1.417) |

Note: ***, **, * are at 1%, 5%, and 10% and the numbers in parentheses are t-statistics.

Table 2 and Figure 4 show the impact of COVID-19 on FinTech lending that precisely measures loan quality by measuring the 90-day success loan settlement rate (TKB90) and the 90-day default rate of loan repayment (TWP90). Before the government announced the COVID-19 pandemic on March 2, 2020, the pre-COVID19 slop of intervention against the 90-day success loan settlement rate showed negative and statistically significant effects at the level of 1%. In contrast, the slop of pre-COVID19 intervention to the 90-day default rate of loan repayment showed a positive impact and was statistically significant at the level of 1%. The amount of value of each slop is negative 0.20 for the 90-day success loan settlement rate and positive 0.2 for the 90-day default rate of loan repayment.

The announcement of the COVID-19 pandemic by the government on March 2, 2020, has impacted the quality of FinTech loans.

252

The impact of COVID19 on the quality of FinTech loans is demonstrated by changes in the 90-day success loan settlement rate and the 90-day default rate of loan repayment. The estimation finds that the impact of COVID19 on the 90-day success loan settlement rate shows negative and statistically insignificant changes in intercepts. In contrast, the impact of COVID19 on the 90-day default rate of loan repayment displays a change in an interception that is marked positive but statistically insignificant. However, the change in intercept contrasts with the trend of post-COVID loan quality intervention (change in slope). As a result, the impact of COVID-19 on the 90-day success loan settlement rate shows a positive trend change, and the 90-day default rate of loan repayment show negative trend change but statistically insignificant.

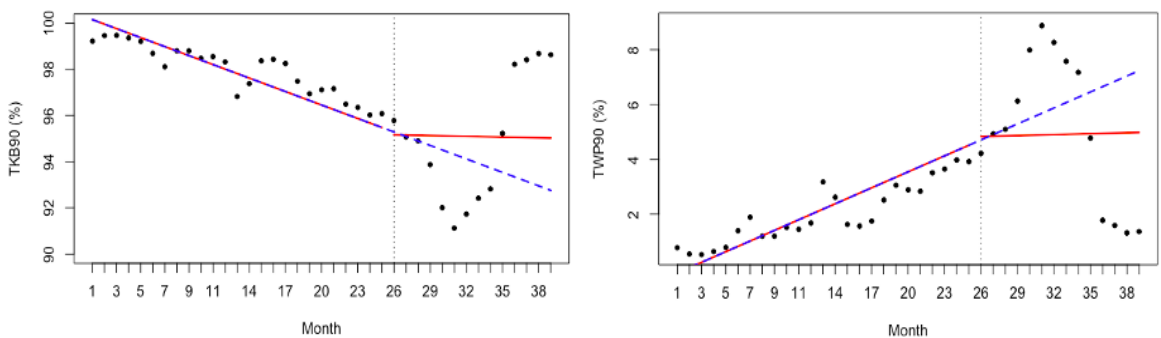

Figure 4 Segmented Regression Model For The Impact of COVID-19 on FinTech Lending Quality Starting in 26th Month (March 2020)

Figure 4 shows the trend of FinTech loan quality pre-COVID19 intervention and post-COVID19 intervention. The pre-COVID19 of the 90-day success loan settlement rate (before March 2, 2020) has decreased, but the post-COVID19 trend (after March 2, 2020) has increased. Meanwhile, the trend of the 90-day default rate of loan repayment has increased pre-COVID19 and decreased post-COVID19. The trend of growing the 90-day success loan settlement rate post-COVID19 above the pre-COVID19

253

trend line, on the contrary, the trend of reducing post-COVID19 of the 90-day default rate of loan repayment below the pre-COVID19 trend line.

Measuring the impact of COVID-19 on FinTech loans is conducted by constructing FinTech counterfactual data and finding the difference to fintech actual data. FinTech counterfactual data discovery uses segmented regression as presented in Table 1 and Table 2 estimation results. In particular, the impact of COVID-19 on FinTech loans can be explored through absolute effect and relative effect in the period after March 2020 starting from April 2020 (2020M04) to April 2021 (2021M04). Specifically, the magnitude of the impact of COVID-19 on FinTech by loan amount and loan quality is shown in Table 3.

Table 3 Impact of COVID-19 on FinTech Lending In Indonesia, 2020M4 - 2021M04

| FinTech lending | | The magnitude of impact of COVID-19 | | | | | | |
|---|---|---|---|---|---|---|---|---|
| | | 2020M04 | 2020M06 | 2020M08 | 2020M10 | 2020M12 | 2021M02 | 2021M04 |
| Δ Borrower from Java | Billion IDR | -3168.79 | -2671.65 | -2174.50 | -1677.36 | -1180.22 | -683.07 | -185.93 |
| | % m-t-m | -48.31 | -37.91 | -28.86 | -20.90 | -13.87 | -7.59 | -1.96 |
| Δ Borrower from Outside Java | Billion IDR | -568.91 | -375.10 | -181.28 | 12.53 | 206.35 | 400.17 | 593.98 |
| | % m-t-m | -52.13 | -31.99 | -14.46 | 0.94 | 14.57 | 26.73 | 37.64 |
| Δ Total Borrower | Billion IDR | -3725.43 | -3042.08 | -2358.73 | -1675.38 | -992.03 | -308.67 | 374.68 |
| | % m-t-m | -48.72 | -37.03 | -26.86 | -17.92 | -10.00 | -2.94 | 3.39 |
| Δ The 90-day success loan settlement rate | Point | 0.06 | 0.43 | 0.80 | 1.16 | 1.53 | 1.90 | 2.27 |
| | % m-t-m | 0.06 | 0.45 | 0.84 | 1.24 | 1.64 | 2.04 | 2.44 |
| Δ The 90-day default rate of loan repayment | Point | -0.06 | -0.43 | -0.79 | -1.16 | -1.52 | -1.89 | -2.25 |
| | % m-t-m | -1.22 | -8.03 | -13.92 | -19.05 | -23.56 | -27.56 | -31.13 |

Note: Δ is the gap between *counterfactual FinTech lending* and *actual FinTech lending*
Billion IDR is Δ *absolute*
% m-t-m is Δ *relative*
Point is Δ *absolute*

The impact of COVID-19 on FinTech transactions varies according to the amount of loan and the lending quality. However, when viewed from absolute changes and relative changes, the amount of negative impact of COVID19 on the number of FinTech lending over time is shrinking or, in other words, changes in slop (post-COVID19 change in slope) between times are increasing. Furthermore, the amount of positive impact of COVID19 on the 90-day success loan settlement rate over time is increasing or, in other words, the post-COVID19 change in slope between times is increasing. Meanwhile, the amount of negative impact of COVID19 on the 90-day default rate of loan repayment over time is getting bigger or, in other words, the post-COVID19 change in slope overtime is decreasing.

The COVID-19 has decreased FinTech lending in Indonesia. The negative impact of COVID-19 is clearly seen on FinTech lending by brorrower from Java. The decline in FinTech lending is due to contracting on real economic activity. The real economic activities that has been declined since the COVID-19 pandemic are tourism, hotel, transportation, trade, and investment sectors (D. A. D. Nasution, Erlina, & Muda, 2020). The negative impact of the COVID-19 pandemic is also seen in Indonesia's manufacturing industry, import exports, and economic growth (Damuri & Hirawan, 2020).

To reduce the negative impact of COVID-19 on FinTech lending in Indonesia, the Financial Services Authority (OJK) has authorized loan restructuring for FinTech peer-to-peer lending players. This policy is set forth in the *Peraturan Otoritas Jasa Keuangan Nomor 58/POJK.05/2020 tentang perubahan atas Peraturan Otoritas Jasa Keuangan Nomor 14/PJOK.05/2020 tentang kebijakan countercyclical dampak penyebaran Coronavirus Disease 2019 bagi Lembaga Jasa Keuangan Non-Bank.* Even so, FinTech lending has not fully undergone a recovery. This condition is due to the enactment of large-scale social restrictions (PSBB) has decreased real economic sector activities related to FinTech lending. Therefore, to expand unbanked populations and MSMEs' access to FinTech lending it is essential to create a new platform business model fintech lending post-COVID19 (Nair, Veeragandham, Pamnani, Prasad, & Guruprasad, 2021).

## CONCLUSION

The paper has discussed the impact of COVID-19 on FinTech lending in Indonesia. The quantifying of the impact of COVID-19 on FinTech lending utilizing interrupted time series (ITS) method on OJK publication FinTech lending data for the period 2018M02 to 2021M04. Empirical facts find variations in the impact of COVID-19 on FinTech lending. In general, COVID-19 negatively impacts FinTech lending. In particular, COVID-19 has a positive effect on the 90-day default rate of loan repayment and a negative to the 90-day success loan settlement rate.

Furthermore, the magnitude of negative impact of post-COVID19 on FinTech lending is gradually shrinking, but the positive impact of post-COVID19 on the 90-day success loan settlement rate is increasing. Furthermore, the negative effect of post-COVID19 on the 90-day default rate of loan repayment is shortening. These findings imply the importance of creating new platform FinTech's post-COVID19 business model to expand public access to peer to peer lending (P2P), especially those untouched by conventional financial institutions.